\documentclass[
    ,final            
  ]
  {aipproc}

\layoutstyle{6x9}

\begin{document}

\title{SN 1987A at the end of its second decade}


\author{Karina Kj\ae r}{
  address={ESO, Karl-Schwarzschild-Strasse 2, D--85478 Garching, Germany}
}

\author{Per Gr\"oningsson}{
  address={Stockholm Observatory, Department of Astronomy, AlbaNova, 
  S--106 91 Stockholm, Sweden}
}

\author{Rubina Kotak}{
  address={ESO, Karl-Schwarzschild-Strasse 2, D--85478 Garching, Germany}
}

\author{Claes Fransson}{
  address={Stockholm Observatory, Department of Astronomy, AlbaNova,
  S--106 91 Stockholm, Sweden}
}

\author{Bruno Leibundgut}{
  address={ESO, Karl-Schwarzschild-Strasse 2, D--85478 Garching, Germany}
}

\begin{abstract}
After nearly two decades at least five emission mechanisms can be found
in SN~1987A. The ejecta continue to glow as a result of the radioactive
decay of long-lived nuclei (mostly $^{44}Ti$), but is fading continuously
because of the expansion and the reduced opacity. The nearly stationary
rings around SN~1987A are still fluorescing from the recombination of
matter originally excited by the soft X-ray emission from the shock
breakout at explosion. The supernova shock reached the inner
circumstellar ring about ten years ago and 
the forward shock is moving through the inner ring
and leaves shocked material behind. This material is excited and
accelerated. The reverse shock illuminates the fast-moving supernova
ejecta as it catches up. And, finally light echoes in nearby
interstellar matter can still be observed. We present here high
resolution spectroscopy in the optical and integral-field spectroscopy
in the near infrared of SN~1987A and its rings. 
\end{abstract}

\maketitle


\section{Introduction}

Monitoring the evolution of a supernova beyond a few years is very rare.
Only a handful of objects have been observable for longer than a year or two
(Leibundgut 1994, Weiler et al. 2002). The clear exception to the rule
is SN~1987A, for which a continuous observational record since its
explosion exists (Arnett et al. 1989, McCray 1993, 2005). It has
been observed in nearly all wavelengths and with the shock now reaching
the circumstellar rings is starting to increase in luminosity again in
the radio (Manchester et al. 2005), X-rays (Park et al. 2006) and
optical/infrared (Bouchet et al. 2006).

The collision of the ejecta of SN~1987A with its circumstellar ring is
creating a series of emission sites in addition to the glow of the
ejecta, the fluorescence of the ring (Fransson et al. 1989, Jakobsen et
al. 1991, Panagia et al. 1991) and light echoes (e.g. Newman \& Rest
2006).  Emission from the ring collision has been observed in the radio
(Gaensler et al. 1997, Manchester et al. 2005), in X-rays (Park et al.
2004, 2005, 2006) and the optical (Michael et al. 2000, 2002, Pun et al.
2002). 

High spatial resolution imaging and high resolution spectroscopy are
the tools of choice for further investigations of SN~1987A. Imaging with
HST continues regularly (Sugerman et al. 2002) and we have been following the
emergence of the emission lines from the shocked material and the
reverse shock with the VLT (Gr\"oningsson et al. 2006, 2007, Kj\ae r et
al. 2007). 

\begin{figure}
  \includegraphics[height=.3\textheight]{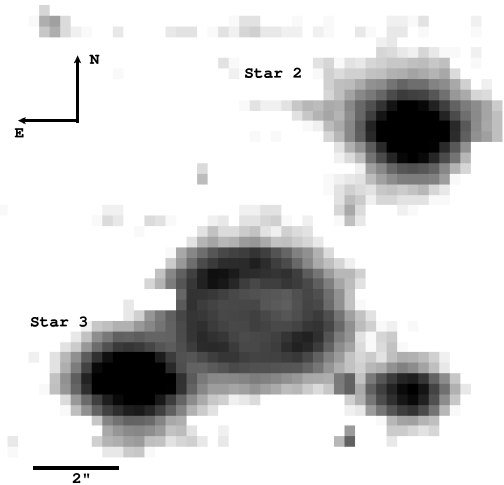}
  \caption{SN~1987A with its inner circumstellar ring as seen in the K-band 
  by SINFONI at the VLT. The nearby unrelated stars 2 and 3 are marked. }
\label{fig:img}
\end{figure}

The extended nature of the ring (and by now the ejecta) limits slit
spectroscopy to individual regions of the ring (cf. Figure~\ref{fig:img}). 
Hence we had to compromise by placing the slit for the high-resolution
spectroscopy across the ejecta and spot 1 (Michael et al. 2000). With
an integral field unit, it is now possible to measure 
the ring and the supernova ejecta without losses and to
reconstruct the ring emission in individual lines (Kj\ae r et al. 2007).
It allows us to cleanly separate the ejecta from the ring as well. 

\section{Coronal emission from the shocked circumstellar ring}

\begin{figure}
\includegraphics[width=\textwidth]{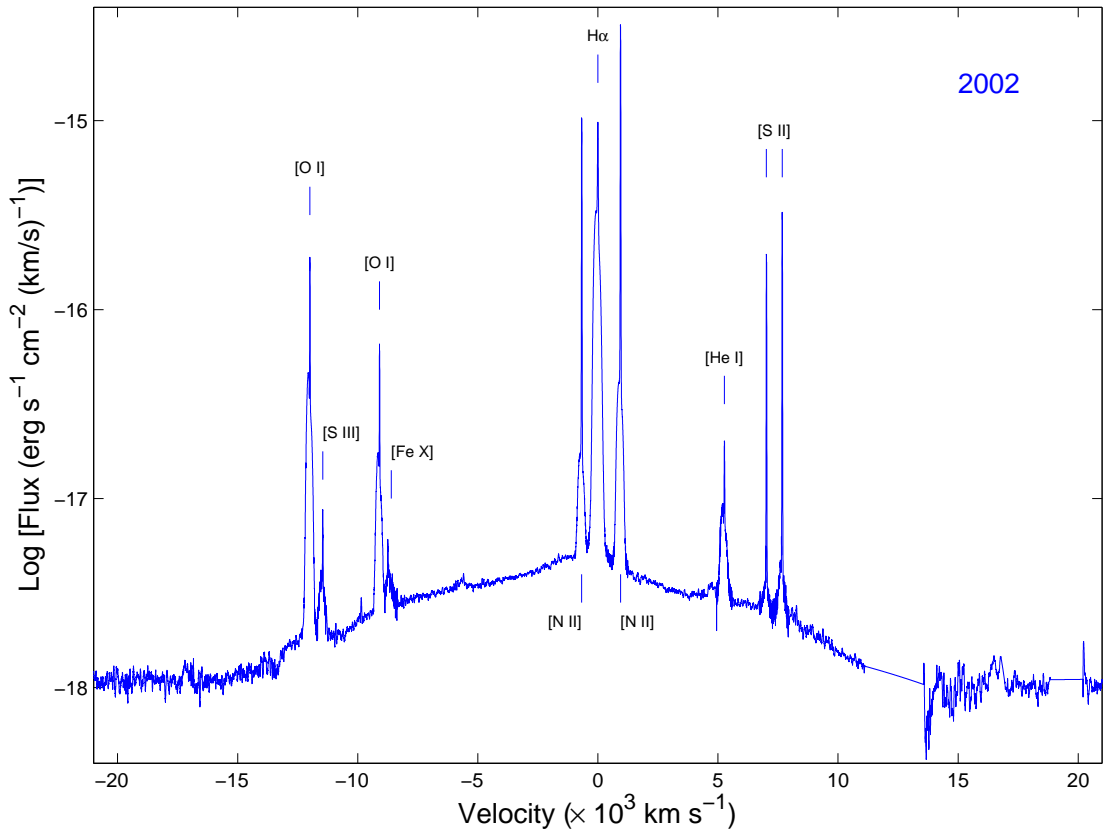} 
\caption{The
 region around H$\alpha$ for SN~1987A 5704 days after explosion. The
 velocity scale is centred on H$\alpha$. Note the very broad component
 originating in the reverse shock, extending to $\sim$~15,000~km~s$^{-1}$. 
 Superimposed on this are the narrow (FWHM~$\sim$~10~km~s$^{-1}$) and 
 intermediate (FWHM~$\sim$~300~km~s$^{-1}$)
 velocity components, coming from the unshocked and shocked ring
 material, respectively, In addition several other low and
 intermediate ionisation lines, as well as the coronal [Fe X], are visible. }
\label{fig:uves_spec}
\end{figure}

We have monitored SN~1987A with a resolution of 40-50,000 since October
1999 using the Ultraviolet and Visual Echelle Spectrograph (UVES) at the
ESO/VLT (Gr\"oningsson et al. 2006, 2007). Fig.~\ref{fig:uves_spec}
shows a spectrum from 2002 around the H$\alpha$ line. The spectrum
consists of three different components described above. The narrow lines
(FWHM~$\sim 10$~km~s$^{-1}$) are from the unshocked, fluorescing
ring material, the broader lines (FWHM~$\sim 250$~km~s$^{-1}$) stem
from the shocked ring and the very wide component (FWZI~$\sim
15000$~km~s$^{-1}$) is the signature of the reverse shock, i.e.  material
in the supernova ejecta. These three components can clearly be
distinguished. 

The reverse shock has been discussed in detail by Michael et al.
(2002), Smith et al. (2005) and Heng et al. (2006). The excitation of
H$\alpha$ is uncertain, but charge exchange is a likely
mechanism. The shock emission continues to increase, but it is
possible that the hydrogen atoms will be ionised by emission from the
forward shock before the reverse shock reaches it and the H$\alpha$
emission from the reverse shock may stop (Smith et al. 2005, Heng et
al. 2006).

\begin{figure}
 \includegraphics[width=\textwidth]{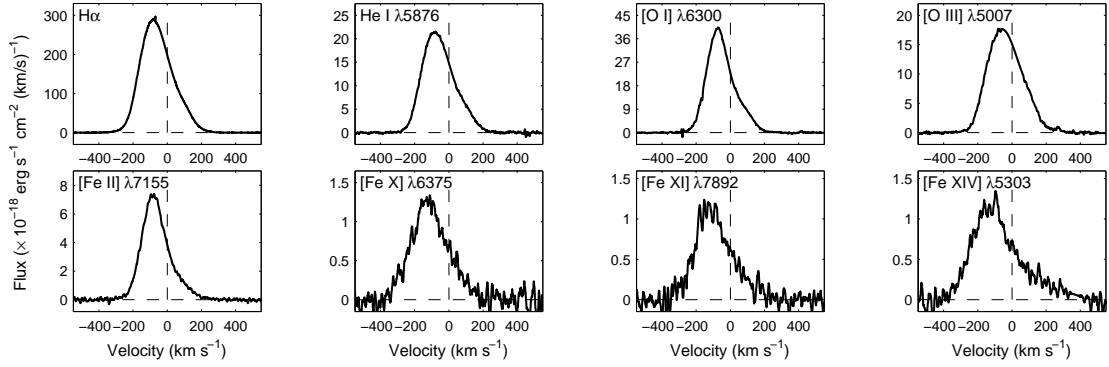}
\caption{Compilation of line profiles from the intermediate velocity
component from day 5702 to 5705 after explosion. The narrow component, as
well as other blends, have been subtracted. Note the larger blue extent
of the coronal lines compared to the low and intermediate ionisation
lines (from Gr\"oningsson et al. 2006).  }
\label{fig:uves_interm}
\end{figure}

In Fig.~\ref{fig:uves_interm} we display a compilation of line profiles
from the intermediate velocity component, where we have removed the
narrow lines as well as blends with other lines.  The lines from the
intermediate velocity component show asymmetric line profiles extending
to FWHM~$\sim 300$ ~km~s$^{-1}$.  There are about 190 lines, among which
roughly 160 from the intermediate-velocity component, in the spectrum
from 310~nm to 1000~nm identified in our spectra (Gr\"oningsson et al.
2007). Several of these lines are coming from highly ionised atoms, in
particular [Fe~X], [Fe~XI], [Fe~XIV], [Ne~V] and [Ar~V]. These arise in
the collisionally ionised gas behind the shock propagating into the
dense protrusions of the ring. In Gr\"oningsson et al. (2006) we
presented a detailed analysis of the coronal lines and derived the
temperature required for their formation. The gas must be heated to
about $2\cdot 10^6$~K, corresponding to a shock velocity of $\sim 350$
~km~s$^{-1}$. The fact that also lines of low ionisation ions, like O
I-III and Fe II, are seen shows that most of the emission is coming from
radiative shocks. We can, however, not exclude that some of the emission
in the coronal lines of FeX--XIV arise from shocks which have not had
time to cool yet. This is indicated by the larger blue extent of these
lines (Fig.~\ref{fig:uves_interm}).

According to the radiative shock model of Gr\"oningsson et al. (2006)
the [Fe~XIII] $\lambda$~1.075 line should be observable in emission.  We
indeed could confirm this line in emission in our ISAAC observations. 

\begin{figure}
  \includegraphics[width=\textwidth]{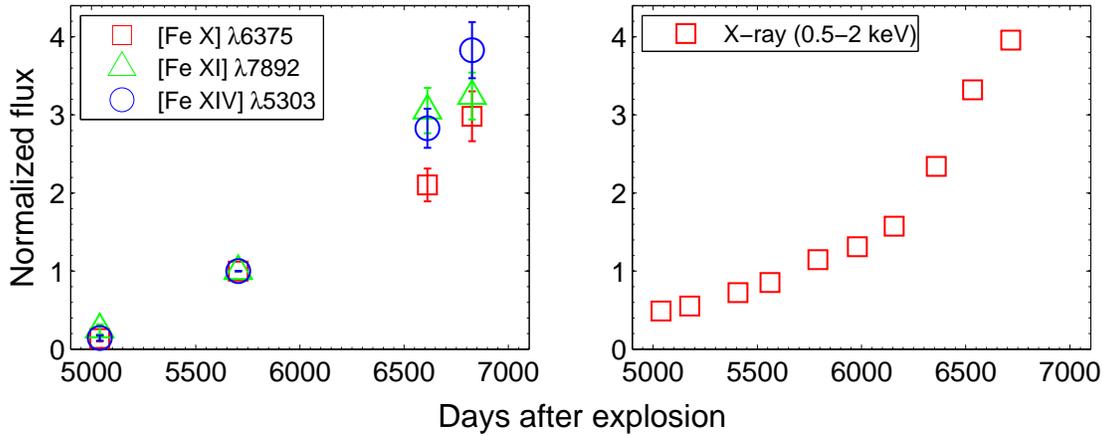}
  \caption{Flux evolution of the coronal lines as measured in optical
  spectroscopy (left) and the flux increase of the X-ray emission from
  SN~1987A (Park et al. 2005). The fluxes have been normalised to the
  ones observed in October 2002 (see Gr\"oningsson et al. 2006 for
  more details).  }
\label{fig:evol}
\end{figure}

A comparison of the evolution of the line fluxes and the X-ray emission
reveals an interesting correlation. The line fluxes of the coronal Fe
lines increased exactly like the soft X-ray (0.5--2 keV) emission as
measured by Chandra (Park et al. 2005; Fig.~\ref{fig:evol}). This,
together with similar shock velocities required to produce the coronal
lines, led Gr\"oningsson et al. (2006) to argue that the coronal lines
form in the same region as the soft X-rays.  The coronal lines therefore
offer a complementary view to the X-rays of the shock interaction. In
this respect the high resolution spectrum of Chandra (Zhekov et al.
2005) is especially interesting, showing lines from, e.g., Fe XVII.

\section{Infrared emission lines}

\begin{figure}
  \includegraphics[height=.5\textheight, angle=-90]{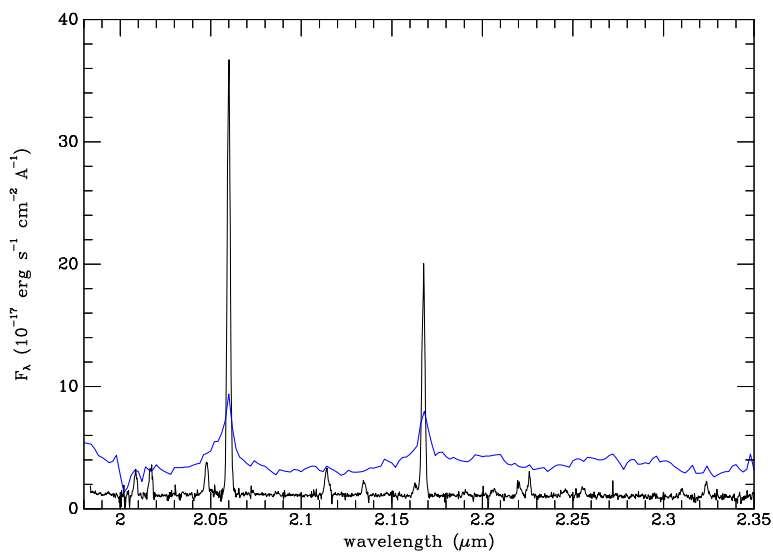}
\caption{K-band spectrum of the integrated ring (lower spectrum) from
November 2004 compared 
to the observation of Fassia et al. (2002) obtained in May 1991 (upper
spectrum). The fluxes have not been adjusted but are as they were observed. 
  }
\label{fig:sinfo_spec}
\end{figure}

Using early Science Verification data of SINFONI, the infrared
integral-field spectrograph at the VLT (Gillesen et al. 2006), we
obtained J, H and K data cubes of SN~1987A and its inner ring 6490
days after explosion. With these data we can determine the integrated
emission of the complete ring or trace the spatial extent of
individual emission lines around the ring. In addition, we can cleanly
separate the ejecta from the ring emission, although the spectral
resolution was not sufficient to separate the narrow and intermediate
components. In Fig.~\ref{fig:sinfo_spec} we show the K-band spectrum
of the integrated ring. He~I $\lambda$2.06 and Br$\gamma$ are the most
prominent lines, but there are several [Fe~II] lines in the near-IR
spectrum (see Kj\ae r et al. 2007 for a detailed list).

Previously, the best NIR spectrum from the AAT (Fassia et al. 2002) is
of much lower spectral resolution than the SINFONI one. The ejecta and
ring contributions could therefore not be resolved. It also is a further
combination of supernova ejecta and circumstellar ring, although it can
be expected that the line emission from the ring did dominate. The
SINFONI spectrum is an integration of the ring emission, excluding the
central part. The two spectra are displayed in
Fig.~\ref{fig:sinfo_spec}.

Due to the higher spectral resolution the narrow He~I and Br$\gamma$
lines appear to have a higher central intensity than some 14 years
before.  Part of this is due to the fact that the earlier spectroscopy
could not resolve the narrow lines. However, as is seen in the optical
UVES spectrum the flux in the intermediate velocity component -- and
both these strong lines should have a contribution from this
component -- has increased in flux. Many other lines have appeared in
the SINFONI spectrum.  Almost all of them are forbidden transitions of
Fe~II. However, we also tentatively identify Na~I, O~I and Ni~II lines
in this spectrum.  The situation in the other IR bands is similar with
several new identifications (Kj\ae r et al. 2007).

\begin{figure}
\includegraphics[height=.4\textheight]{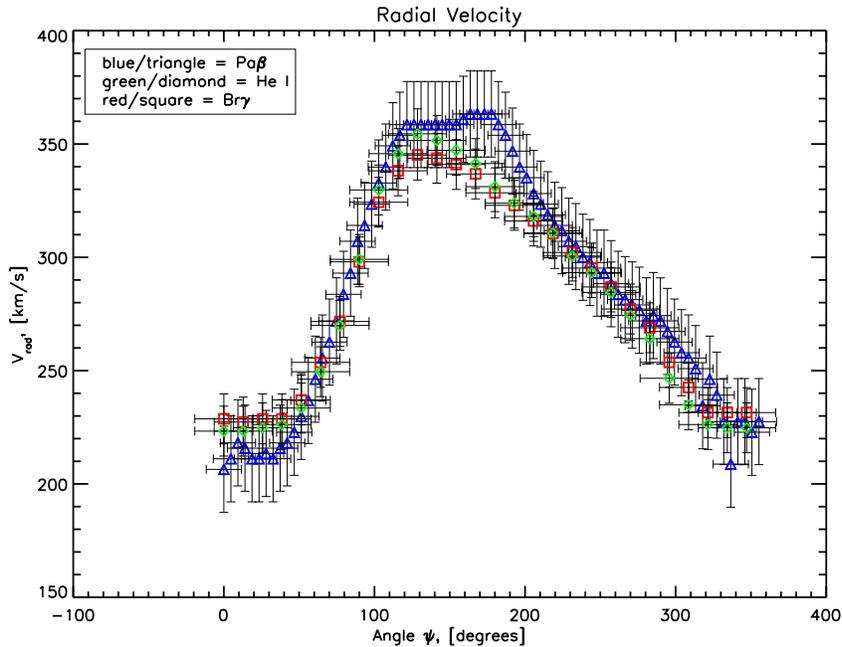}
  \caption{The radial velocity as measured around the ring. The
  angle is defined with $\psi(North)=0^\circ$ and
  $\psi(East)=90^\circ$. Spot~1, the first interaction spot to show up
  in HST imaging (e.g. Pun et al. 2002) is located at $\psi=29^\circ$
  (Sugerman et al. 2002).
  }
\label{fig:v_rad}
\end{figure}

The spectral resolution of SINFONI is $\sim 150$~km~s$^{-1}$ and hence we
cannot distinguish between the fluorescing gas and the shocked matter.
Nevertheless, we are able to trace the velocity dispersion and the
radial velocity around the ring. Figure~\ref{fig:v_rad} shows the
azimuthal velocity distribution for the three prominent lines in the ring
(Pa$\beta$, Br$\gamma$ and He~I $\lambda$2.06). 

It is now possible to determine the rest frame velocity of the ring
assuming that the acceleration around the ring has been constant. This
is clearly not the case, as the flat-topped velocity distribution of
Pa$\beta$ shows. There is a clear deviation from the expected sine curve
for a tilted ring. The reason is, of course, that the density of the
ring protrusions and the shock interaction are not the same around the
ring. The eastern part has had the first interaction and hence a higher
shock velocity is expected in these regions. Also, the ring is not
simply inclined perpendicular to the line of sight but also rotated out
of the plane of the sky with the eastern sector more distant than the
western part.  We derive from the velocity curve a rotation angle of
$\sim$ 10$^\circ$, consistent with the offset seen in the major axis of
the ring on the sky (Jakobsen et al. 1991, Panagia et al.  1991).
Nevertheless assuming that the asymmetries average out over the complete
ring, we can find the systemic velocity of the ring, and presumably of
SN~1987A itself. A first, preliminary analysis yields
$v_{SN1987A}=283\pm8$~km~s$^{-1}$, which is consistent with the value
derived from the high-resolution UVES spectroscopy, $286.7 \pm 0.1$~km~s$^{-1}$. 

\section{The ejecta spectrum}

The UVES spectra clearly show a broad H$\alpha$ line extending to
$\sim 15,000$~km~s$^{-1}$ (Fig.~\ref{fig:uves_spec}). This is
discussed in detail by Smith et al. (2005) and Heng et al. (2006), who
show that this originates from the reverse shock, propagating back
into the ejecta. In addition, there is also a component in this
line, as well as in [Ca~II] $\lambda 7324$, coming from the radioactively
excited core.

Detection of emission from the supernova ejecta are more difficult in
the SINFONI spectrum. The only line we securely can identify as coming
from the ejecta is [Fe~II] $\lambda$1.64. It has a velocity width of
about 4000~km~s$^{-1}$, which corresponds to the nickel core of
SN~1987A.  Unfortunately, the early science verification data do not
have sufficient spatial resolution to allow us to investigate the
shape of the ejecta. However, there are strong indications from HST
observations that the supernova ejecta is deviating strongly from a
spherical distribution (Wang et al. 2002). With deeper spectroscopy and adaptive optics
supported observations, we hope to obtain a clearer picture of the
ejecta in the coming years. 

\section{Conclusions}

It is clear that with the kinematic information we will be able to
further investigate the evolution of the supernova shock as it envelopes the
circumstellar ring in the coming years. SN~1987A is living up to being
a true teenager. It is transforming itself into a mature supernova
remnant, but on the way has to go through some turbulent times as the
supernova shock overtakes the circumstellar ring. SN~1987A represents
a fascinating laboratory to study how a supernova interacts with its
surrounding. 

We continue our spectroscopic monitoring of the emission in the
optical and near-infrared. Integral-field spectroscopy is an ideal
tool to investigate the many emission sites. 


\end{document}